\documentclass[10pt]{article}
\usepackage{graphicx}
\usepackage[T1]{fontenc}
\usepackage[utf8]{inputenc}
\usepackage{amssymb}
\usepackage{amsfonts}
\usepackage{dsfont}
\usepackage{mathtools}
\usepackage{amsthm}
\usepackage{amsmath}
\usepackage{relsize}
\usepackage{textcomp}
\usepackage{eurosym}
\usepackage{stmaryrd}
\usepackage{xcolor}
\usepackage[multiple]{footmisc}
\usepackage{bigints}
\usepackage{geometry}
\begin{document}

\title{Dealing with multi-currency inventory risk in FX cash markets}

\author{Alexander \textsc{Barzykin}\footnote{HSBC, 8 Canada Square, Canary Wharf, London E14 5HQ, United Kingdom, alexander.barzykin@hsbc.com.} \and Philippe \textsc{Bergault}\footnote{\'Ecole Polytechnique, CMAP, 91128 Palaiseau, France, philippe.bergault@polytechnique.edu.} \and Olivier \textsc{Guéant}\footnote{Université Paris 1 Panthéon-Sorbonne, Centre d'Economie de la Sorbonne, 106 Boulevard de l'Hôpital, 75642 Paris Cedex 13, France, olivier.gueant@univ-paris1.fr.}}
\date{}

\maketitle
\setlength\parindent{0pt}

\begin{abstract}

In FX cash markets, market makers provide liquidity to clients for a wide variety of currency pairs.
Because of flow uncertainty and market volatility, they face inventory risk. 
To mitigate this risk, they typically skew their prices to attract or divert the flow and trade with their peers on the dealer-to-dealer segment of the market for hedging purposes. 
This paper offers a mathematical framework to FX dealers willing to maximize their expected profit while controlling their inventory risk. 
Approximation techniques are proposed which make the framework scalable to any number of currency pairs.

\vspace{5mm}

\end{abstract}

\section*{Introduction}

As all market makers, FX dealers are naturally portfolio managers. By providing liquidity to clients in multiple currency pairs, they build inventory and have to manage the ensuing inventory risk which is a subtle combination of uncertainty of the client flow, market liquidity and market price risk at the portfolio level.\\

The management of inventory risk in financial markets has been a topic of recent interest in the academic field of quantitative finance, starting with the seminal paper  \cite{avellaneda2008high} by Avellaneda\footnote{Marco Avellaneda passed away while we were finishing this paper. This paper is therefore a natural opportunity to pay him a tribute for his major contributions to the field and beyond.} and Stoikov who revived an old economic literature on the topic that dated back to the 1980s (see, for instance, Ho and Stoll \cite{ho1981optimal}). 
Avellaneda and Stoikov paved the way to a long list of contributions. In a nutshell,\footnote{See the books \cite{cartea2015algorithmic} and \cite{gueant2016financial} for a detailed bibliography.} Guéant et al. provided in \cite{gueant2013dealing} a detailed analysis of the stochastic optimal control problem introduced in \cite{avellaneda2008high} and proposed closed-form approximations of the optimal quotes. New features were then progressively added to get closer and closer to reality: several trade sizes, client tiering, risk externalization through hedging, etc. 
Cartea and Jaimungal, along with various coauthors, replaced in \cite{cartea2014buy} the original expected utility framework of \cite{avellaneda2008high} by a more intuitive one closer to mean-variance optimization and added new features while studying the impact of parameters ambiguity in a series of papers. 
\\

To be used in practice, market making models need to take into account the dependence structure between assets since market makers or market making algorithms typically cover dozens or hundreds of assets. 
A mathematical framework for multi-asset market making has been proposed in \cite{gueant2016financial, gueant2017optimal}, but computing the optimal quotes almost always requires to solve differential equations in very high dimension (the number of equations typically grows exponentially with the number of assets). 
Several techniques have been proposed to tackle the curse of dimensionality: the use of a few numbers of risk factors, the use of neural networks and reinforcement learning techniques, and the use of a reduction technique toward a linear-quadratic control problem (see \cite{bergault2021closed}) that provides surprisingly good approximations of the optimal market making strategy.\\

Existing multi-asset market making models consider assets labelled in the same currency. This is typically consistent with the problem faced by market makers in most asset classes (e.g. corporate bonds in Europe or in the US). However, the problem faced by FX dealers is different: each currency pair provides indeed a valuation of one currency in terms of the other. The inventory of a given currency is usually managed in the most liquid, so-called direct currency pair, typically against USD.
However, non-USD pairs, so-called crosses, are also very important. The presence of crosses introduces both complexity and opportunities since there are several ways to achieve the same result (e.g. buying EURGBP is equivalent to buying EURUSD and selling GBPUSD). Our model is, to our knowledge, the first to address the problem faced by an FX dealer who quotes a wide variety of currency pairs, including crosses, and can mitigate inventory risk by hedging on dealer-to-dealer (D2D) and/or all-to-all platforms. In particular it answers subtle questions where liquidity and correlation issues are intertwined, e.g. for positively correlated EURUSD and GBPUSD legs, what should be the spread of EURGBP? How should it compare to the sum of leg spreads? How to attract offsetting client flow or deter risky client flow in an optimal way when the marker maker is active in both direct pairs and crosses? How to optimally hedge the portfolio when trading platforms exist for direct pairs and some crosses?\\

As a consequence of FX market specific characteristics, our model differs from classical market making models in many ways, going from price dynamics to how trade sizes are defined, to the choice of the mathematically relevant state variables. 
To obtain optimal quotes and optimal hedging rates, one needs to solve a high-dimensional differential equation, but the techniques proposed in \cite{bergault2021closed} can be adapted so that approximations of the optimal strategies can be obtained by solving a low-dimensional matrix Riccati-like differential equation (the dimensionality of the differential equations to solve only grows quadratically with the number of currencies, hence linearly with the maximal number of currency pairs).\\

We start by presenting our multi-currency market making model and show that, through a smart choice of the state variables, 
the problem boils down to solving a partial differential equation.
We then demonstrate how the ideas developed in \cite{bergault2021closed} can be adapted to approximate the true value function in closed form up to the computation of the solution of a matrix Riccati-like differential equation. We hereby claim that our approach is sufficiently scalable for practical use since the size of the (square) matrix mentioned above corresponds to the number of currencies. We further discuss  the relevance of our closed-form approximations and illustrate numerically the resulting market making strategy in a market with 5 major currencies: USD, EUR, JPY, GBP and CHF.
In particular, we demonstrate how correlated currency pairs are engaged through pricing and hedging when managing the risk in a single pair and show the impact of the presence of several currency pairs on the internalization vs. externalization dilemma faced by FX dealers \cite{butz2019internalisation}.
With the help of Monte Carlo simulations, we confirm that the inventory probability distribution is shaped up by the portfolio risk profile and that the observed risk autocorrelation time is much shorter than any of the currency pair position autocorrelation times, leading to cost savings.
\\

\section*{Multi-currency market making model}

\subsection*{Description of the framework}

\paragraph{The market and the dealer}

We consider a market with $d$ currencies and an FX dealer in this market over a time interval of length $T$. The dealer trades currency pairs through two channels: with clients at prices streamed electronically, and with other dealers on the D2D segment of the market or on all-to-all platforms.

\paragraph{Exchange rates}

In order to model exchange rates, we consider throughout this paper that currency~$1$ (which typically stands for USD) is the reference currency and we introduce $d$ stochastic processes denoted by $(S^1_t)_{t\in [0,T]}, \ldots, (S^d_t)_{t\in [0,T]}$ that model the market price of each of the $d$ currencies in terms of the reference one (in particular,  $S^1_t = 1$ for all $t\in [0,T]$). Then, the market exchange rate of any currency pair can be deduced from the market exchange rates of the two currencies with respect to currency $1$.

\paragraph{Streamed pricing ladders and clients demand for currency pairs}

The dealer divides their clients into $N$ tiers and streams them pricing ladders at the bid and at the ask for each currency pair, i.e. exchange rates for each side and different sizes. Mathematically, we introduce for each couple $(i,j) \in \{1, \ldots, d\}^2$ and each tier $n \in \{1, \ldots, N\}$ a $\mathbb R_+^* - $marked point process $J^{n,i,j}(dt,dz)$ modelling transactions with clients from tier $n$ regarding the currency pair $(i,j)$, where $z$ is the size variable measured in reference currency. Formally, if $J^{n,i,j}(dt,dz)$ has a jump corresponding to size $z$ at time $t$, it means that the market maker sells to the client a ``quantity'' $z/S^j_t$ of currency $j$ and receives in exchange a payment in currency $i$ in line with the corresponding streamed pricing ladder. To build our model, we assume that this payment is decomposed into a ``quantity'' $z/S^i_t$ of currency $i$ and fees, denoted by $z\delta^{n,i,j}(t,z)$, that are accumulated on a separate account labeled in reference currency -- hence $\delta^{n,i,j}$ represents the markup (possibly negative) in percentage or basis points. The process $J^{n,i,j}(dt,dz)$ has an intensity kernel $(\nu^{n,i,j}_t(dz))_{t\in [0,T]}$ verifying
$$\nu^{n,i,j}_t(dz) = \Lambda^{n,i,j}\left(z, \delta^{n,i,j} (t,z)\right)dz,$$
where $\Lambda^{n,i,j}$ is called the intensity function of the process $J^{n,i,j}(dt,dz)$. Following \cite{barzykin2021market}, we assume that this function $\Lambda^{n,i,j}$ -- which models the probability of transaction per unit of time as a function of the demanded size and proposed quote -- is of the logistic type:\footnote{Generalizations are of course straightforward.}
$$\Lambda^{n,i,j}(z,\delta)=\lambda^{n,i,j}(z) f^{n,i,j}(z,\delta) \quad \text{with} \quad f^{n,i,j}(z,\delta) =  \frac{1}{1+e^{\alpha^{n,i,j}(z) + \beta^{n,i,j}(z) \delta}}.$$

\paragraph{D2D segment}

In addition to skewing quotes (internalization) to attract or divert client flow, the FX dealer can trade currency $i$ against currency $j$ on the D2D segment of the market (externalization). To model this form of hedging we introduce for each couple $(i,j) \in \{1, \ldots, d\}^2$ with $i<j$ a process $\left(\xi^{i,j}_t \right)_{t\in [0,T]}$ which models the amount (expressed in reference currency) per unit of time of currency $i$ bought by the dealer and paid in currency $j$. Unlike what happened for pricing, we only consider couples $(i,j)$ with $i<j$: $\xi^{i,j}_t$ can be negative if the dealer buys currency $j$ and pays in currency $i$. When trading at rate $\xi^{i,j}_t$, we assume that the dealer incurs execution costs\footnote{In reality, not all pairs are available for trading on D2D platforms. This would correspond to very high execution costs in our model.} modeled by a term $L^{i,j}(\xi^{i,j}_t)$ (accounted in dollars like the above fees), where the function $L^{i,j}$ is chosen of the form $L^{i,j}(\xi) = \psi^{i,j} |\xi| + \eta^{i,j} |\xi|^{1+\phi^{i,j}}$ ($\phi^{i,j} = 1$ throughout this paper).

\subsection*{Dynamics of the state variables}

The dealer has inventories in $d$ currencies modeled by $d$ processes $(q^1_t)_{t\in [0,T]}, \ldots, (q^d_t)_{t\in [0,T]}$. Wrapping up the above, we get that the dynamics of inventories is given by 
$$\forall i\in \{1, \ldots, d\}, \quad dq^i_t = \sum_{n=1}^N \underset{j\neq i}{\sum_{ j=1}^d} \int_{z \in \mathbb R_+^*} \frac z{S^i_t} \left(J^{n,i,j}(dt,dz) -  J^{n,j,i}(dt,dz) \right)+ \left( \sum_{j=i+1}^d \frac{\xi^{i,j}_t}{S^i_t} - \sum_{j=1}^{i-1}\frac{\xi^{j,i}_t}{S^i_t} \right)dt.$$
Moreover, the dynamics of the account where fees and execution costs are accumulated is
$$dX_t = \sum_{n=1}^N \sum_{1\le i\neq j \le d} \int_{z \in \mathbb R_+^*} z\delta^{n,i,j} (t,z) J^{n,i,j}(dt,dz) - \sum_{1\le i<j\le d} L^{i,j}\left(\xi^{i,j}_t \right).$$

Regarding the dynamics of market exchange rates, we assume that
$$\forall i\in \{1, \ldots, d\}, \quad dS^i_t = \mu^i_t S^i_t dt + \sigma^i S^i_t dW^i_t + k^i \left( \sum_{j=i+1}^d \xi^{i,j}_t - \sum_{j=1}^{i-1}\xi^{j,i}_t \right) S^{i}_t dt,$$
where $(\mu^i_t)_{t\in [0,T]}$ is a deterministic drift, $\sigma^i \ge 0$ is the volatility of currency $i$ with respect to the reference currency, $k^i$ is a linear permanent market impact parameter,\footnote{We only consider the impact of our trades on the D2D segment of the market. Adverse selection could also be modeled through an impact of transactions with clients on the market exchange rates. Modeling adverse selection in a realistic and implementable way is in fact an active field of research.} and $\left(W^1_t,\ldots,W^d_t \right)_{t\in [0,T]}$ is a $d$-dimensional correlated Brownian motion. Of course, $\mu^1 = \sigma^1 = k^1 = 0$.\\

It is convenient for what follows to use vector and matrix notations:
$$S_t = \left(S^1_t, \ldots, S^d_t \right)^\intercal \in \mathbb R^d, \quad \mu(t) = \mu_t = \left(\mu^1_t, \mu^2_t, \ldots, \mu^d_t \right)^\intercal \in \mathbb R^d \quad  \text{and} \quad \Sigma = (\rho^{i,j} \sigma^i \sigma^j)_{1\le i, j \le d} \in \mathcal S^+_{d}(\mathbb R),$$ where $\rho^{i,j} = \frac{d\langle W^i,W^j\rangle}{dt}$.\\

Given this dynamics for prices, we conclude that for all $i\in \{1, \ldots, d\}$, the process $\left(Y^i_t \right)_{t\in [0,T]}$ =  $\left(q^i_t S^i_t \right)_{t\in [0,T]}$ corresponding to the inventory of currency $i$ measured in reference currency has the following Markovian dynamics:
\begin{align}
dY^i_t &= \mu^i_t Y^i_{t-} dt + \sigma^i Y^i_{t-} dW^i_t + k^i \left( \sum_{j=i+1}^d \xi^{i,j}_t - \sum_{j=1}^{i-1}\xi^{j,i}_t \right) Y^{i}_{t-} dt\nonumber\\
&\quad + \sum_{n=1}^N \underset{j\neq i}{\sum_{ j=1}^d} \int_{z \in \mathbb R_+^*} z \left(J^{n,i,j}(dt,dz) -  J^{n,j,i}(dt,dz) \right) + \left( \sum_{j=i+1}^d \xi^{i,j}_t - \sum_{j=1}^{i-1}\xi^{j,i}_t \right)dt.\nonumber    
\end{align}
In what follows, we denote by $(Y_t)_{t\in [0,T]}$ the vector of inventories measured in reference currency, i.e. $Y_t= \left(Y^1_t, \ldots, Y^d_t \right)^\intercal \in \mathbb R^d$.\\

\subsection*{Optimization problem}

The FX dealer wants to maximize the Mark-to-Market value of their portfolio at time $T$, while mitigating inventory risk. Mathematically, we assume that they want to maximize
$$\mathbb E \left[ X_T + \sum_{i=1}^d Y^i_T- \frac{\gamma}{2} \int_0^T Y_t^\intercal \Sigma Y_t dt - \ell \left(Y_T \right) \right]$$
over the admissible pricing $(\delta^{n,i,j})_{1\le n \le N, 1\le i\neq j \le d}$ and execution $(\xi^{i,j})_{1\le i<j \le d}$ controls, where $\gamma >0$ represents the risk aversion of the market maker and $\ell$ is a penalty for the remaining inventory at time $T$.\footnote{$\ell$ could account for the market impact when unwinding.}\\

\section*{Towards optimal quotes}

\subsection*{Hamilton-Jacobi-Bellman equation and optimal quotes}

Applying Ito's formula to the process $\left(X_t + \sum_{i=1}^d Y^i_t  \right)_{t \in [0, T]}$ allows us to see that the above optimization problem is equivalent to maximizing
\begin{align*}
\mathbb{E}&\Bigg[\int\limits_{0}^{T} \Bigg\lbrace \sum_{n=1}^N \sum_{1\le i\neq j \le d} \int_{z \in \mathbb R_+^*} \Big(z\delta^{n,i,j}(t,z)  \Lambda^{n,i,j}(z,\delta^{n,i,j}(t,z))  \Big)dz + \sum_{i=1}^d \bigg(\mu^i_t + k^i \Big( \sum_{j=i+1}^d \xi^{i,j}_t - \sum_{j=1}^{i-1}\xi^{j,i}_t \Big) \bigg)Y^{i}_t\\
&\qquad \qquad - \sum_{1\le i<j\le d} L^{i,j}\left(\xi^{i,j}_t \right) - \frac{\gamma}{2} Y_t^\intercal \Sigma Y_t  \Bigg\rbrace dt - \ell(Y_T) \Bigg].\nonumber
\end{align*}

This optimization problem can be addressed with the tools of stochastic optimal control. We denote by $\theta:[0,T]\times \mathbb R^d \rightarrow \mathbb{R}$ the value function of this stochastic control problem. The associated Hamilton-Jacobi-Bellman equation is
\begin{equation}
\begin{cases}
 \!&0 = \partial_t \theta(t,y) + y^\intercal \mu(t) + y^\intercal \mathcal D(\mu(t)) \nabla_y\theta(t,y) - \frac{\gamma}{2}y^\intercal \Sigma y + \frac 12\text{Tr}\left(\mathcal D (y)\Sigma \mathcal D (y) D^2_{yy} \theta(t,y)\right)\\
\!& \qquad + \text{\scalebox{0.6}[1]{$\bigint$}}_{\!\!\mathbb{R}_{+}^{*}} \underset{n=1}{\overset{N}{\mathlarger \sum}} \underset{1\le i\neq j \le d}{{\mathlarger \sum}} zH^{n,i,j} \left(z,\frac{\theta(t,y) -  \theta(t,y+ze^i - z e^j) }{z}\right)\lambda^{n,i,j}(z) dz\\
\!& \qquad + \underset{1\le i<j \le d}{{\mathlarger \sum}} \mathcal H ^{i,j} \left(\partial_{y^i}\theta(t,y) - \partial_{y^j}\theta(t,y) + k^i y^i \left(1 + \partial_{y^i}\theta(t,y) \right) -  k^j y^j \left(1 + \partial_{y^j}\theta(t,y) \right)\right),\\
\!&\theta(T,y) = -\ell(y),
\end{cases}
\label{eqn:HJB}
\end{equation}
where
\begin{equation}
H^{n,i,j}:(z,p)\in\mathbb R_+^* \times \mathbb{R} \mapsto \underset{\delta }{\sup}\ f^{n,i,j}(z,\delta)(\delta-p),\nonumber
\end{equation}
\begin{equation}
\mathcal H^{i,j}:p\in\mathbb{R} \mapsto \underset{\xi}{\sup}\ p\xi - L^{i,j}(\xi),\nonumber
\end{equation}
and $\mathcal D (y)$ denotes the diagonal $d\times d$ matrix such that $\mathcal D (y)_{i,i} = y_i$ for all $i\in\{1, \ldots, d\}.$\\

It is proved in \cite{gueant2017optimal} that for all $(n,i,j)$, the supremum in the definition of $H^{n,i,j}(z,p)$ is reached at a unique $\bar \delta^{n,i,j}(z,p) = (f^{n,i,j})^{-1} \left(-\partial_p{H^{n,i,j}} (z,p)  \right)$ and this function can easily be computed numerically in the logistic case we consider. If $\theta$ is known, we therefore obtain the optimal quotes in the following form
\begin{align}\label{optquotes}
 \delta^{n,i,j*}(t,z) = \bar \delta^{n,i,j} \left(z, \frac{\theta(t,Y_{t-}) -  \theta(t,Y_{t-}+ze^i - z e^j) }{z} \right).   
\end{align}

Similarly, the optimal trading rates are given by 
\begin{align}\label{optrates}
\xi^{i,j*}_t = {\mathcal H^{i,j}}' \left(\partial_{y^i}\theta(t,Y_{t-}) - \partial_{y^j}\theta(t,Y_{t-}) + k^i Y^i_{t-} \left(1 + \partial_{y^i}\theta(t,Y_{t-}) \right) -  k^j Y^j_{t-} \left(1 + \partial_{y^j}\theta(t,Y_{t-}) \right) \right).   
\end{align}

\subsection*{Approximation of the value function and optimal quotes}

Following the same ideas as in \cite{bergault2021closed}, we now approximate for $n \in \{1, \ldots, N\}$ and for each couple $(i,j) \in \{1, \ldots, d\}^2$ with $i\neq j$ the Hamiltonian function $H^{n,i,j}$ by a quadratic function $\check H^{n,i,j}$:
$$\check H^{n,i,j}(z,p) = \alpha_0^{n,i,j}(z) + \alpha_1^{n,i,j}(z) p + \frac 12 \alpha_2^{n,i,j}(z) p^2,$$
where a natural choice is of course
$$\alpha_0^{n,i,j}(z) =  H^{n,i,j}(z,0),\quad \alpha_1^{n,i,j} = \partial_p{H^{n,i,j}}(z,0),\quad \text{and} \quad \alpha_2^{n,i,j} = \partial^2_{pp}{H^{n,i,j}}(z,0).$$

The structure of the problem leads us to approximate the Hamiltonian terms associated with $\mathcal H^{i,j}$ by $0$ since $\mathcal H^{i,j}$ is typically flat around $0$ when $\psi^{i,j} > 0$.\\

We then consider the new equation
\begin{equation}
\begin{cases}
 \!&0 = \partial_t \check \theta(t,y) + y^\intercal \mu(t) + y^\intercal \mathcal D(\mu(t)) \nabla_y \check\theta(t,y) - \frac{\gamma}{2}y^\intercal \Sigma y + \frac 12\text{Tr}\left(\mathcal D (y)\Sigma \mathcal D (y) D^2_{yy} \check \theta(t,y)\right)\\
\!& \qquad + \text{\scalebox{0.6}[1]{$\bigint$}}_{\!\!\mathbb{R}_{+}^{*}} \underset{n=1}{\overset{N}{\mathlarger \sum}} \underset{1\le i\neq j \le d}{{\mathlarger \sum}} z\check H^{n,i,j} \left(z,\frac{\check\theta(t,y) -  \check \theta(t,y+ze^i - z e^j) }{z}\right)\lambda^{n,i,j}(z) dz,\\
\!&\check \theta(T,y) = -\ell(y).
\end{cases}
\label{eqn:HJBap0}
\end{equation}

If $\ell(y) = y^\intercal \kappa y$ with $\kappa$ a semi-definite positive symmetric matrix, then Eq. \eqref{eqn:HJBap0} has a solution of the form $\check \theta(t,y) = -y^\intercal A(t)y - y^\intercal B(t) - C(t)$ where $t \mapsto A(t) \in \mathcal{S}_d$, $t \mapsto B(t) \in \mathbb R^d$ and $t \mapsto C(t) \in \mathbb R$ solve differential equations. As the value of $C$ is irrelevant for what follows, we only report here the equations for $A$ and $B$:
\begin{align}\label{ODEsys}
\begin{cases}
A'(t) &= 2A(t) M A(t) - \Sigma \odot A(t) - 2\mathcal D(\mu(t)) A(t)  - \frac{\gamma}{2}\Sigma\\
B'(t) &= \mu(t) -  \mathcal D(\mu(t)) B(t) +  2A(t) V + 2A(t) \tilde V\left(A(t) \right) + 2A(t) M B(t),\\
A(T) &=\kappa ,\quad B(T) = 0,
\end{cases}
\end{align}
where $\odot$ denotes the Hadamard (i.e. element-wise) product,
$$M = \mathcal D \left(\left(\overline M + \overline M^\intercal \right)  U \right) - \left(\overline M + \overline M^\intercal \right), \quad V = \left(\underline M - \underline M^\intercal \right)U, \quad \text{and} \quad  \tilde V (A) = \left(\overline V(A) - \overline V(A)^\intercal \right)U,$$
with $U = (1,\ldots, 1) ^\intercal \in \mathbb R^d$, $\overline M$ and $\underline M$ two $d\times d$ matrices such that
$$\overline M_{i,j} = \sum_{n=1}^N  \int_{\mathbb R_+^*}\alpha_2^{n,i,j}(z) z \lambda^{n,i,j}(z)dz \quad \text{and} \quad \underline M_{i,j} = \sum_{n=1}^N  \int_{\mathbb R_+^*}\alpha_1^{n,i,j}(z) z \lambda^{n,i,j}(z)dz,$$
and
$$\overline V(A)= \overline{\mathcal D}(A) P + P \overline{\mathcal D}(A) -2 P \odot A \quad \text{with} \quad P_{i,j} = \sum_{n=1}^N  \int_{\mathbb R_+^*}\alpha_2^{n,i,j}(z) z^2 \lambda^{n,i,j}(z)dz,$$
where $\overline{\mathcal D}(A)$ is a $d\times d$ diagonal matrix with the same diagonal as $A$.\\

The ODE system \eqref{ODEsys} involves a matrix Riccati-like differential equation whose solution can be approximated very easily using an Euler scheme (i.e. without suffering from the curse of dimensionality, unlike when using an inventory grid as in the case of the original Hamilton-Jacobi-Bellman equation). Once $A$ and $B$ are obtained, approximations of the optimal strategies can be computed by replacing $\theta$ by $\check \theta$ in Eqs. \eqref{optquotes} and~\eqref{optrates}. We thereby obtain 
$$\check \delta^{n,i,j}(t,z) =  \bar \delta^{n,i,j} \bigg(z,\Big(\big(2Y_{t-} + z (e^i-e^j)\big)^\intercal A(t) + B(t)^\intercal  \Big) (e^i-e^j)  \bigg),$$
and
\begin{eqnarray*}
\check \xi^{i,j}_t = {\mathcal H^{i,j}}' \Big(-\big(A(t) Y_{t-} + B(t)\big)^\intercal(e^i-e^j)&\!\!+\!\!& k^i Y^i_{t-} \left(1 -\big(A(t) Y_{t-} + B(t)\big)^\intercal e^i \right)\\ &\!\!-\!\!&  k^j Y^j_{t-} \left(1 -\big(A(t) Y_{t-} + B(t)\big)^\intercal e^j \right) \Big).
\end{eqnarray*}

\section*{Numerical results and discussion}

Before illustrating the market making strategy proposed above, it is noteworthy that we have validated, when $d=2$, our approximations against the optimal strategy computed thanks to the solution of Eq. \eqref{eqn:HJB} approximated with a monotone implicit Euler scheme on an inventory grid. Using parameters inspired from earlier work \cite{barzykin2021algorithmic, barzykin2021market}, we studied both the strategies and the corresponding efficient frontier. Only under extreme conditions which are not practically relevant, such as strong order flow asymmetry (as high as fivefold) and very high or very low risk aversion, have we detected significant deviations.\\

For our illustrations, we consider a market with 5 major currencies: USD, EUR, JPY, GBP and CHF. We considered two tiers and a discretization of trade sizes corresponding to $1$, $5$, $10$, $20$ and $50$ M\$ for all currency pairs.  We used the parameters documented in Table~\ref{parameters_table} unless specified otherwise. These parameters have been selected by analysing a subset of HSBC market making franchise, as previously described in~\cite{barzykin2021market}. However, they should not be considered as representative of HSBC but rather of a typical FX dealer. Standard currency pair naming convention is respected except that we used CHFUSD and JPYUSD instead of USDCHF and USDJPY to be consistent with our model (USD being the reference currency in our examples).\\

In what follows, both the drift vector $\mu$ and the terminal penalty $\kappa$ are assumed to be $0$. The time horizon is set to $T = 0.05$ days (72 minutes) that ensures convergence towards stationary quotes and hedging strategy at time $t = 0$. In particular, to compute the market making strategy, we used $A(0)$ and $B(0)$ instead of $A(t)$ and $B(t)$ throughout to mimic what would happen in the stationary case.\\

\begin{figure}[!h]
\centering
\includegraphics[width=0.83\textwidth]{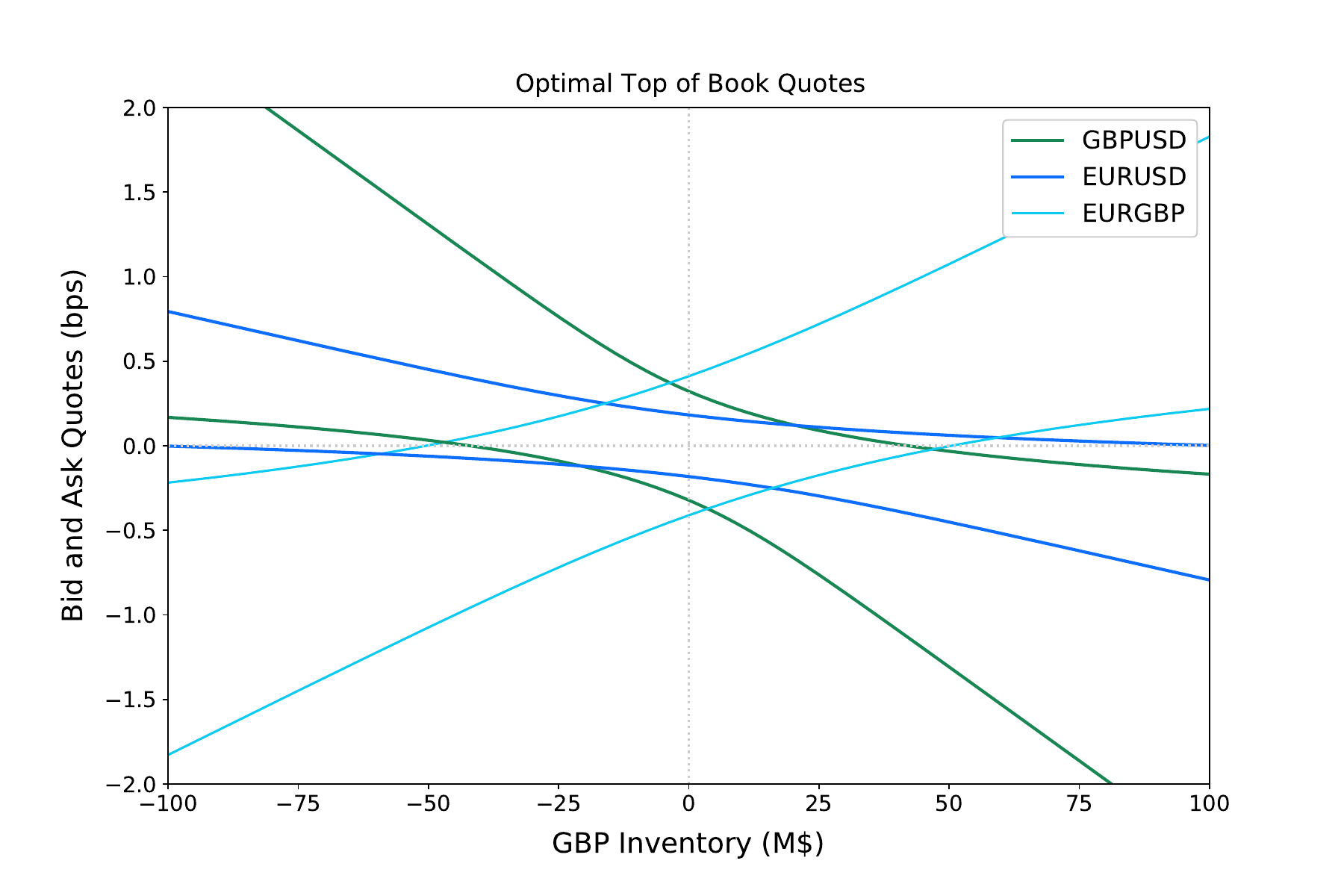}
\caption{Optimal top of book pricing for the currency pairs EURUSD, GBPUSD and EURGBP as functions of GBP inventory with other inventories set to $0$ (tier 1). The curves represent respectively $\delta^{1,X,Y}$ and $-\delta^{1,Y,X}$ for each currency pair XY. Risk aversion: $\gamma = 20$ (M\$)$^{-1}$. 
}
\label{optimal_tob_majors}
\end{figure}

Fig.~\ref{optimal_tob_majors} illustrates top of book pricing (i.e. for a size of $1$M\$) of EURUSD, GBPUSD and EURGBP as functions of GBP inventory, keeping the other inventories at $0$, for tier 1. GBPUSD pricing looks familiar, with a skew to attract risk-offsetting flow and divert risky flow. Without correlation and without the cross, EURUSD pricing would be unaffected by GBP inventory. Here, instead, positive correlation leads to a protective pricing strategy for EURUSD. The pricing of the cross pair EURGBP also attracts or diverts the flow as a function of the GBP inventory, and this in turn influences the pricing of EURUSD.

\begin{table}[!h]
\begin{center}
\vspace{10pt}
{\bf Direct pairs} \\
\begin{tabular}{|lccccccc|}
\hline
Pair & $\sigma \left(\frac{\text{bps}}{\sqrt{\text{day}}}\right)$
& $\lambda(z) \left(\frac{1}{\text{day}}\right)$ 
& $\alpha$ & $\beta \left(\frac{1}{\text{bps}}\right)$ 
& $\psi$ (bps) 
& $\eta \left(\frac{\text{bps}\cdot\text{day}}{\text{M\$}}\right)$ 
& $k \left(\frac{\text{bps}}{\text{M\$}}\right)$ \\
\hline
EURUSD & 80 & 900, 540, 234, 90, 36  & -1.9, -0.3 & 11, 3.5 & 0.1 & $10^{-5}$ & $5\cdot10^{-3}$ \\
GBPUSD & 70 & 600, 200, 150, 40, 10 & -1.4, 0.0 & 5.5, 2.0 & 0.15 & $1.5\cdot10^{-5}$ & $7\cdot10^{-3}$ \\
CHFUSD & 60 & 420, 140, 105, 28, 7 & -1.2, 0.0 & 4.5, 1.9 & 0.25 & $2.5\cdot10^{-5}$ & $8\cdot10^{-3}$ \\
JPYUSD & 60 & 825, 375, 180, 105, 15 & -1.6, -0.1 & 9.0, 3.0 & 0.1 & $1.5\cdot10^{-5}$ & $6\cdot10^{-3}$ \\
\hline
\end{tabular}

\vspace{10pt}
{\bf Crosses} \\
\begin{tabular}{|lcccccc|}
\hline
Pair & $\rho$
& $\lambda(z) \left(\frac{1}{\text{day}}\right)$ 
& $\alpha$ & $\beta \left(\frac{1}{\text{bps}}\right)$ 
 & $\psi$ (bps) 
& $\eta \left(\frac{\text{bps}\cdot\text{day}}{\text{M\$}}\right)$ \\
\hline
EURGBP & 0.6 & 400, 50, 25, 20, 5 & -0.5, 0.5 & 3.5, 2.5 & 0.25 & $3\cdot10^{-5}$ \\
EURCHF & 0.5 & 400, 50, 25, 20, 5 & -0.5, 0.5 & 3.5, 2.5 &  0.25 & $3\cdot10^{-5}$ \\
EURJPY & 0.3 & 400, 50, 25, 20, 5 & -0.5, 0.5 & 3.5, 2.5  & 0.25 & $3\cdot10^{-5}$ \\
GBPCHF & 0.3 & 160, 20, 10, 8, 2 & -0.5, 0.5 & 3.5, 2.5 & 0.4 & $5\cdot10^{-5}$ \\
GBPJPY & 0.2 & 160, 20, 10, 8, 2 & -0.5, 0.5 & 3.5, 2.5 & 0.4 & $5\cdot10^{-5}$ \\
CHFJPY & 0.4 & 80, 10, 5, 4, 1 & -0.5, 0.5 & 3.5, 2.5 & 0.4 & $5\cdot10^{-5}$ \\
\hline
\end{tabular}
\caption{
Parameters for the currency pairs. The correlation coefficient $\rho$ provided for crosses describes correlation between the corresponding dollar-based legs.
Size ladders are the same for all pairs in reference currency, i.e. $z = 1, 5, 10, 20, 50$ M\$.
Two client tiers with different $\alpha$ and $\beta$ parameters (independent of $z$) are considered for each pair. Intensity amplitudes $\lambda(z)$ are taken to be the same for each tier.
}
\label{parameters_table}
\end{center}
\end{table}

\begin{figure}[!h]
\centering
\includegraphics[width=0.85\textwidth]{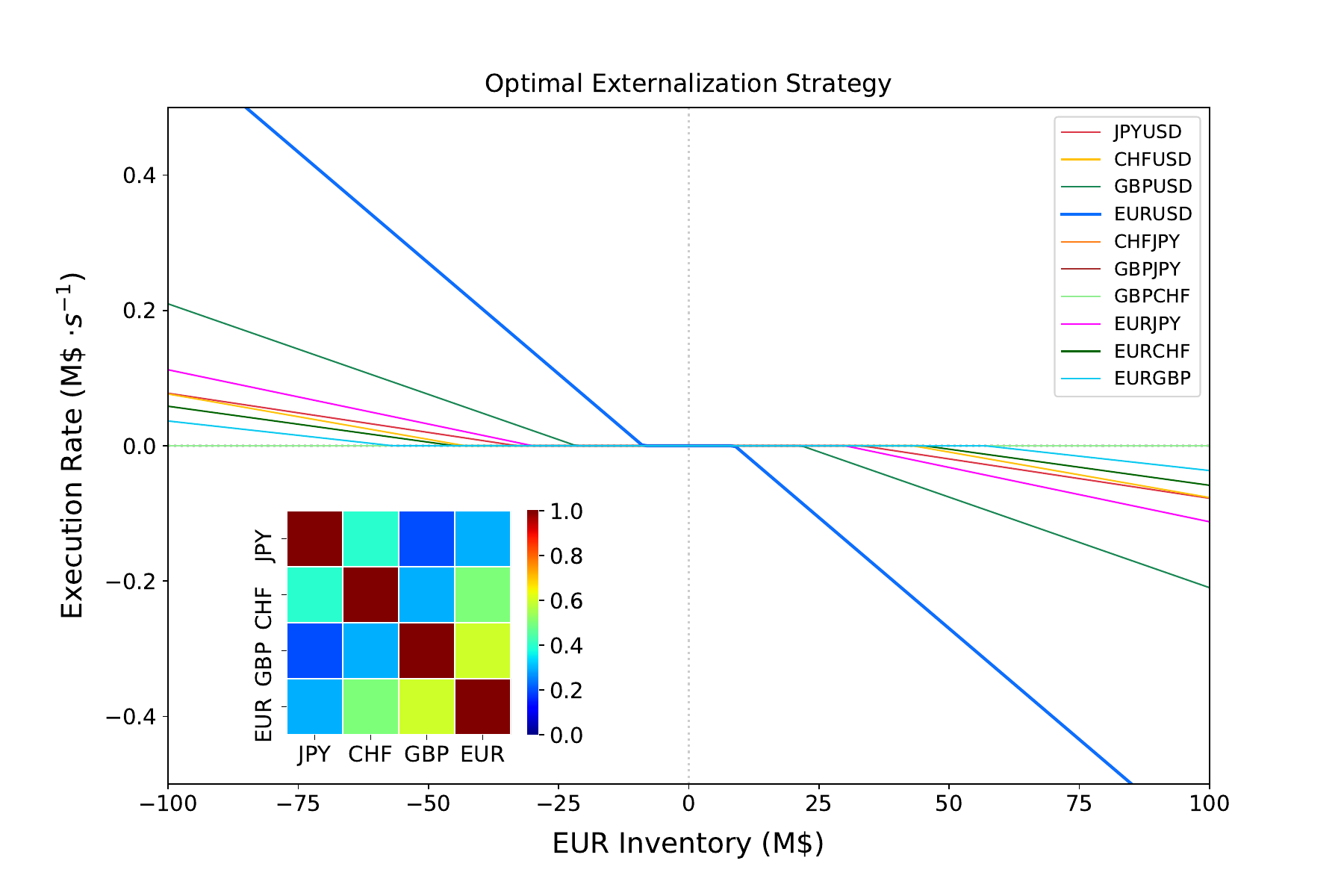}\\
\caption{Optimal execution rates for the different currency pairs as functions of EUR inventory with other inventories set to $0$. Inset: correlation matrix.
Risk aversion: $\gamma = 20$ (M\$)$^{-1}$.
}
\label{optimal_execution_majors}
\end{figure}

Fig.~\ref{optimal_execution_majors} shows optimal hedging strategy as a function of EUR inventory when other inventories are equal to  zero. EURUSD execution rate displays a familiar pattern with pure internalization zone in the middle and nearly linear growth for larger positions. Understandably, when the inventory becomes very large the dealer may want to offload part of the risk into other correlated direct currency pairs and crosses.
The main reason is that for large inventories the price skew has likely already been exploited and one cannot expect much different client flow when skewing further.\\

\begin{figure}[!h]
\centering
\includegraphics[width=0.68\textwidth]{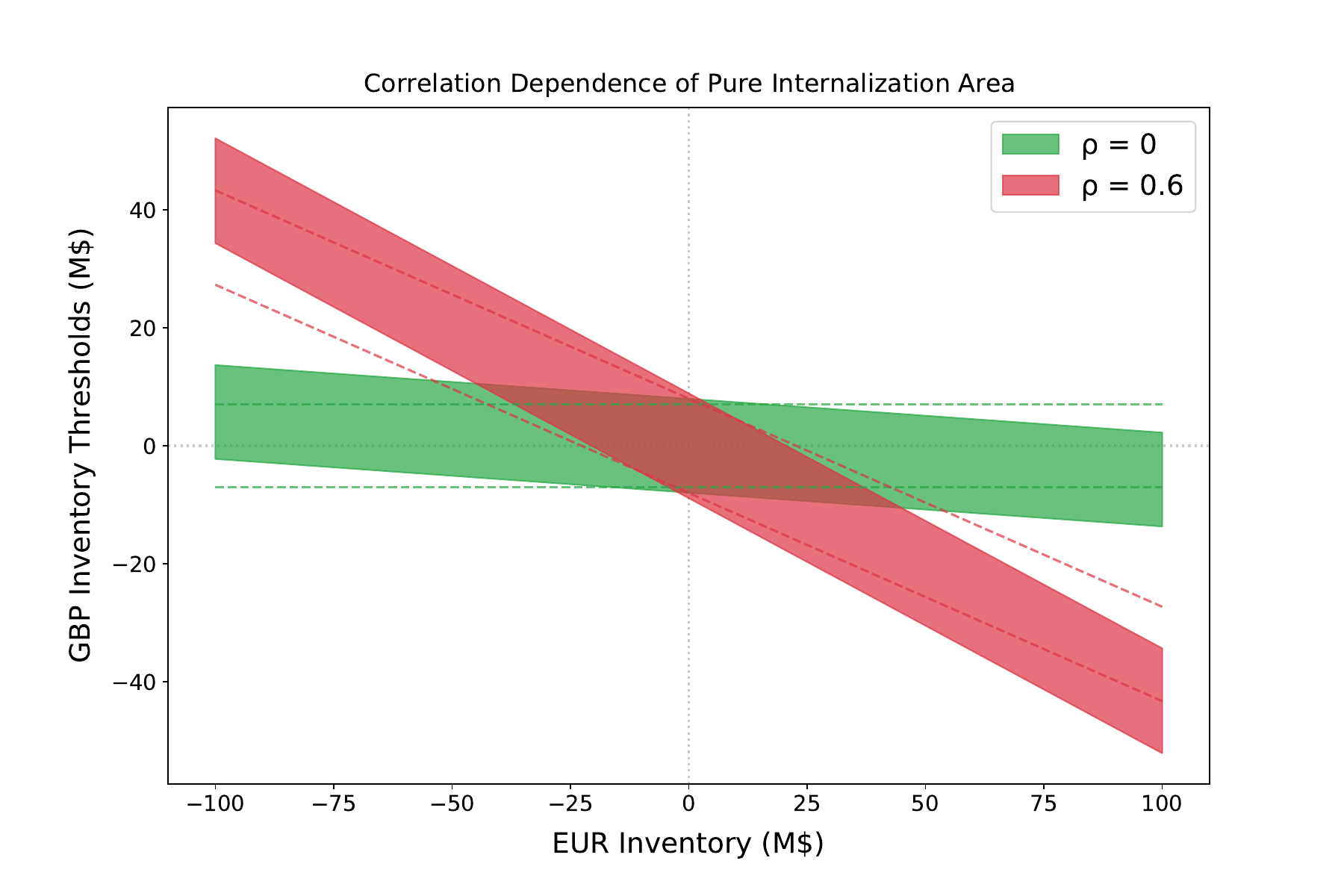}\\
\caption{Pure internalization zone thresholds for GBP as functions of EUR inventory in a market with USD, EUR and GBP. Different correlation levels are color coded as labeled. Dashed line correspond to a market without the cross, i.e. only \mbox{EURUSD} and \mbox{GBPUSD}. Risk aversion: $\gamma = 20$ (M\$)$^{-1}$.
}
\label{thresholds_trio}
\end{figure}

Fig.~\ref{thresholds_trio} explores the effect of correlation and the influence of the cross pair on the pure internalization zone in the case of the three currencies USD, EUR and GBP. In line with intuition, the pure internalization zone is slanted because of correlation: it is not always optimal for a dealer to start hedging externally when positions in correlated currencies already mitigate part of the risk. The pure internalization zone is even more slanted in the presence of the cross pair which can attract client flow on its own and limit the necessity to hedge externally. Interestingly, the presence of the cross influences the internalization zone even without correlation: it is slanted and not horizontal even when $\rho = 0$.\\

\begin{figure}[h!]
\centering
\includegraphics[width=0.68\textwidth]{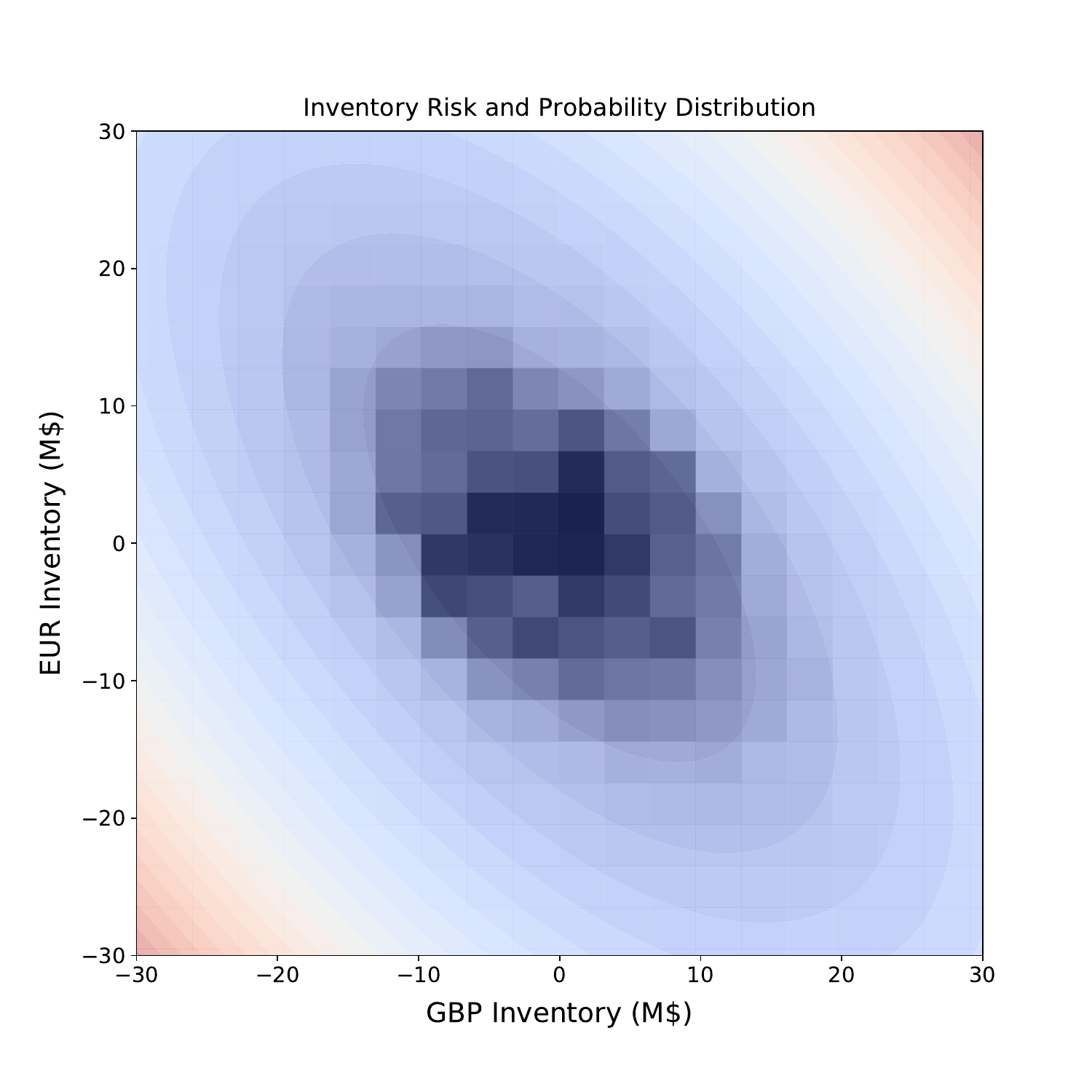}\\
\caption{Inventory risk (countour plot) defined as $\frac{\gamma}{2} y^\intercal \Sigma y$ and inventory probability distribution associated with simulations of the market making strategy 
(2d histogram on the basis of a $10^6$ second long Monte Carlo trajectory) in a market with EURUSD, GBPUSD and EURGBP.
Risk aversion: $\gamma = 20$ (M\$)$^{-1}$.
}
\label{inventory_pdf_trio}
\end{figure}

Once the optimal strategy has been computed, one can follow \cite{barzykin2021market} and simulate the inventories resulting from the use of the market making strategy via standard Monte Carlo procedure. Fig.~\ref{inventory_pdf_trio} illustrates the probability distribution of inventories in EUR and GBP in the case of a market with the three currencies USD, EUR and GBP. This empirical distribution is superimposed onto the risk contour plot and we clearly see that risk drives the inventory distribution.\\

\begin{figure}[!h]
\centering
\includegraphics[width=0.85\textwidth]{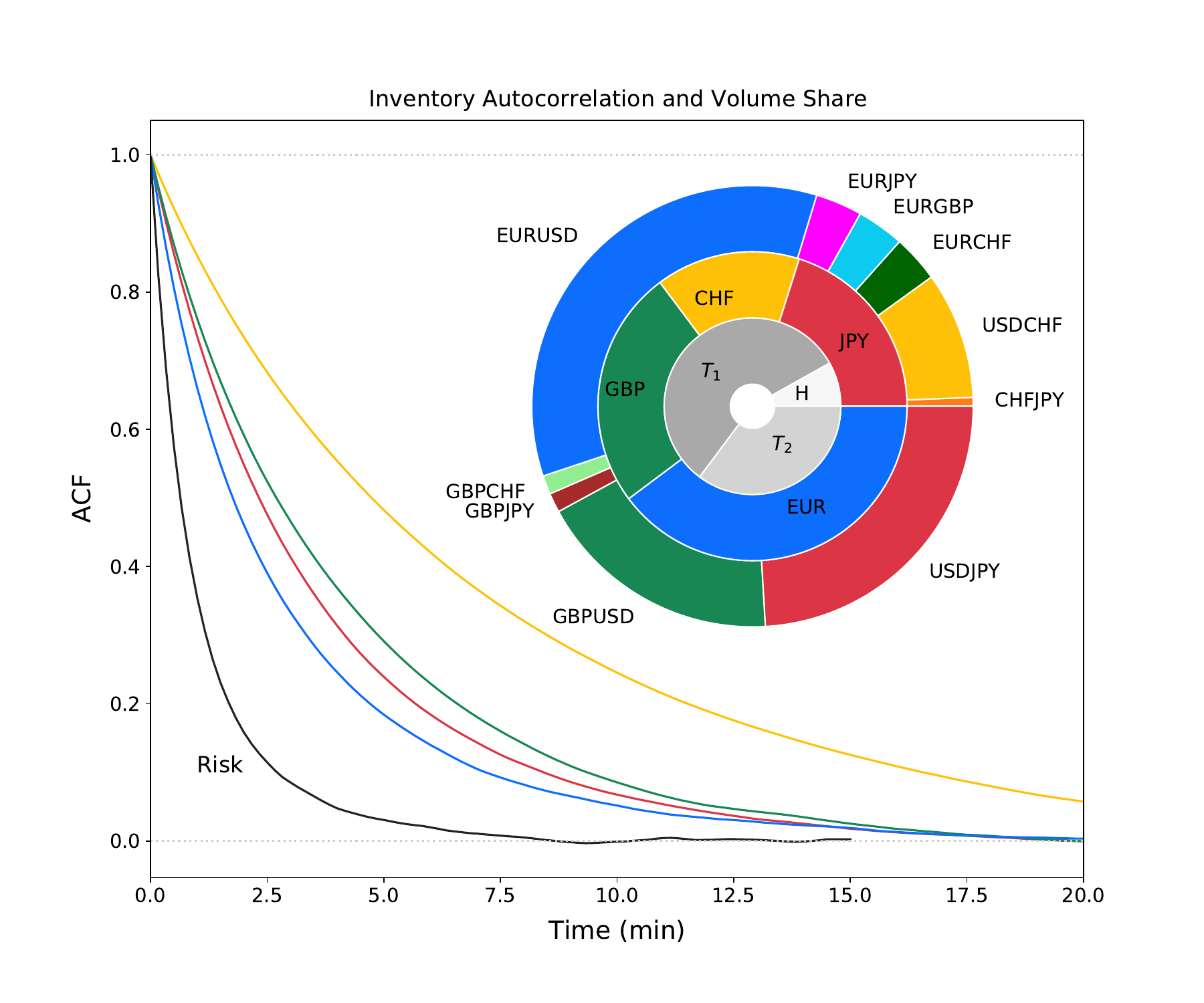}\\
\caption{Component inventory and portfolio risk autocorrelation functions on the basis of a $5 \cdot 10^6$ second long Monte Carlo trajectory. Inset pie chart shows the corresponding volume share by currency pair (outer layer), 
P\&L share by currency outside of USD (middle layer) and traded volume distribution among client tiers and hedging 
(inner layer, $H$ denotes external hedging, $T_1$ and $T_2$ stand for the tiers). Risk aversion: $\gamma = 20$ (M\$)$^{-1}$.
}
\label{inventory_acf_majors}
\end{figure}

Fig.~\ref{inventory_acf_majors} returns to the case of a market with the 5 currencies and shows that the inventory autocorrelation functions of individual pairs decay much slower than the risk autocorrelation function. This means that the dealer is able to offload risk fast while essentially trading slower and thus saving on impact and transaction cost. The figure also provides information on volume share, P\&L share and internalization ratio.
The latter is consistent with previously reported typical internalization levels of about 80\% for G10 currencies by top-tier banks \cite{schrimpf2019fx}.\\

\newpage

\section*{Concluding remarks}

We have introduced and analyzed in detail a multi-currency market making model incorporating fundamental risk controls taking into account correlations, client tiering, pricing ladders and external hedging with transaction costs and market impact. Approximation techniques are proposed which make the framework scalable to any number of currency pairs and thus offering immediate practical application to the FX industry. 
The results obtained demonstrate efficient risk reduction due to optimization at portfolio rather than individual currency pair level.

\section*{Statement and acknowledgment}

The results presented in this paper are part of the research works carried out within the HSBC FX Research Initiative. 
The views expressed are those of the authors and do not necessarily reflect the views or the practices at HSBC. 
The authors are grateful to Richard Anthony (HSBC) for helpful discussions and support throughout the project.

\bibliographystyle{plain}

\end{document}